\documentclass[aps,prl,showpacs,preprintnumbers,nofootinbib,twocolumn]{revtex4}
\usepackage{bm}
\usepackage{latexsym}
\usepackage{dcolumn}
\usepackage{amsfonts,amssymb}
\usepackage{graphicx,epsfig}
\usepackage{psfrag}
\usepackage{amsmath,amssymb}

\def\be {\begin{equation}}
\def\ee {\end{equation}}
\def\ba {\begin{eqnarray}}
\def\ea {\end{eqnarray}}
\def\nn {\nonumber}
\def\a  {\alpha}
\def\b  {\beta}
\def\c  {\gamma}

\def\d  {\delta}

\def\e  {\epsilon}

\def\k  {\kappa}
\def\L  {\Lambda}

\def\n  {\nu}

\def\O  {\Omega}
\def\p  {\pi}
\def\la {\label}
\def\le {\left}
\def\ri {\right}

\def\f {\frac}

\def\no {\noindent}
\def\bi {\begin{itemize}}
\def\ei {\end{itemize}}

\def\laq{\hbox{~}\raise 0.4ex\hbox{$<$}\kern -0.8em\lower 0.62ex\hbox{$\sim$}\hbox{~}}
\def\gaq{\hbox{~}\raise 0.4ex\hbox{$>$}\kern -0.7em\lower
  0.62ex\hbox{$\sim$}\hbox{~}}

\def\beq{\begin{equation}}
\def\eeq{\end{equation}}
\def\br{\begin{eqnarray}}
\def\er{\end{eqnarray}}
\def\benu{\begin{enumerate}}
\def\eenu{\end{enumerate}}
\def\nn{\nonumber} 

\def\l{\left}
\def\r{\right}    

\begin{document}
\title{How robust is the entanglement entropy - area relation?}

\author{Saurya Das$^{\S}$ and S. Shankaranarayanan$^{\P}$}
\affiliation{$^{\S}$ Dept. of Physics, University of Lethbridge,
4401 University Drive, Lethbridge, Alberta T1K 3M4, Canada}
\email{saurya.das@uleth.ca}
\affiliation{$^{\P}$ HEP Group, International Centre 
for Theoretical Physics, Strada costiera 11, 34100 Trieste, Italy}
\email{shanki@ictp.trieste.it}

\begin{abstract}
We revisit the problem of finding the entanglement entropy of a scalar
field on a lattice by tracing over its degrees of freedom inside a
sphere. It is known that this entropy satisfies the area law --
entropy proportional to the area of the sphere -- when the field is
assumed to be in its ground state. We show that the area law continues
to hold when the scalar field degrees of freedom are in generic
coherent states and a class of squeezed states. However, 
when excited states are considered, the entropy scales as a lower 
power of the area. This suggests that for large horizons, the ground state entropy
dominates, whereas entropy due to excited states gives power law
corrections. We
discuss possible implications of this result to black hole entropy.
\end{abstract}

\maketitle

Although classical black holes (BHs) have infinite entropy and zero
temperature, Bekenstein -- inspired by the area increase theorem of
general relativity -- proposed that BHs have entropy proportional to
horizon area $A_H$. This, together with Hawking's discovery that BHs
radiate with the temperature $T_H= \hbar c^3/(8\pi G M)$ have given rise
to the Bekenstein-Hawking area law for BH entropy:
%
\beq
S_{BH} = \frac{A_H}{4 \ell_P^2 } ;\qquad
\ell_{_P} \equiv \sqrt{\frac{\hbar G}{c^3}}~~ (\mbox{Planck length}).
\eeq

The area (as opposed to volume) proportionality of BH entropy has been
an intriguing issue for decades. Attempts to understand this problem can
be broadly classified into two classes: (i) those that count
fundamental states such as $D$-Branes and spin-networks, which are
supposed to model BHs \cite{stringsetc}, and (ii) those that
study entanglement entropy \cite{bkls,sred} and its variants such as
the brick-wall model and Shakarov's induced gravity \cite{thooft}.

In the case of entanglement entropy, which is of interest in this
work, it is assumed that the von Neumann entropy 
%
\beq
S = - Tr[\rho \ln(\rho)]
\label{eq:def-vNE}
\eeq
of quantum fields due to correlations between the exterior and
interior of the BH horizon, accounts for black hole entropy. Such
correlations imply that the state of field, when restricted outside
the horizon, is mixed, although the full state may be pure
\cite{bkls,sred}.
Although this entropy is ultra-violet divergent, a suitable short
distance cut-off [${\cal O}(\ell_{_P})$] gives $S \propto A_H$ (it was
argued in \cite{bkls} that this entropy must be formally divergent).
This idea gained further credence, when it was shown that even for
{\it Minkowski space-time} (MST), tracing over the degrees of
freedom inside a hypothetical sphere (of radius $R$), gives rise to
the entropy of the form $0.3~(R/a)^2$ where $a$ is the lattice spacing
\cite{bkls,sred}
(it was shown in \cite{arom} that quantum fluctuations inside the 
sub-volume scale as its bounding area as well). 
Thus, the area-law may be a direct consequence of
entanglement alone.

However, a crucial assumption was made in the analyses of \cite{bkls}
and \cite{sred}, that all the harmonic oscillators (HOs) -- resulting
from the descretization of the scalar field - are in their ground
state (GS). Thus the natural question which one would ask is: How
sensitive is the {\it area law} to the choice of the quantum state of
the HOs?

In a recent paper, the current authors had investigated this problem
for a simpler system of two coupled oscillators, and found two
interesting results \cite{ads}: (i) the entropy remains unchanged if
the GS oscillator wave functions are replaced by generalized coherent
states (GCS),
and (ii) the entropy could increase substantially (as
much as $50\%$) even if one of the oscillators is in its first excited
state (ES). This raises the possibility that for the more interesting
case of $N$-coupled oscillators (modeling a free scalar field),
deviations from the area law could result if excited states are taken
into account.  We address this issue in this work. When the
oscillators are in GCS and a class of 
squeezed states (SS), we show analytically that entanglement entropy
exactly equals that of the ground state, implying that the area law
remains valid.
For ESs, of the form of superpositions of a number of wave-functions, each of
which has exactly one HO in the first ES, we show numerically that
the entanglement entropy still scales as a power of the area, but that 
the power is now less than unity. The more terms there are in the
superposition, the less is this power \cite{PI}. 

%

The Hamiltonian for a free scalar field ($\varphi$) is
%
\beq
H = \f{1}{2} \int d^3x \le[ \p^2(x) + |\vec\nabla\varphi(\vec x) |^2\ri] 
\eeq
where $\pi$ is the momentum conjugate of $\varphi$. Decomposing
$\varphi$ and $\pi$ in terms of real spherical harmonics $(Z_{lm})$ i. e.,
$$ \varphi_{lm} (r) [\p_{lm}(r)] = r \int d\O~Z_{lm} (\theta,\phi)
\varphi (\vec r) [\p (\vec r)] \, , $$
and discretizing on a radial lattice $r \rightarrow r_j \, ,$ [with 
$r_{j+1} - r_j = a = M^{-1}$ and $L=(N+1)a$ is the box size], we get, 
%
\br
H_{lm}& = & \f{1}{2a} \sum_{j=1}^N
\le[ \p_{lm,j}^2 + \le(j+\f{1}{2}\ri)^2
\le( \f{\varphi_{lm,j}}{j}
- \f{\varphi_{lm,j+1}}{j+1}
\ri)^2 \r. \nn \\
&+& \l. \f{l(l+1)}{j^2}~\varphi_{lm,j}^2 
\ri]  \quad , \quad H= \sum_{lm} H_{lm} \, ,
\label{disc1}
\er
%
\noindent \hspace*{-10pt} where {\small $\varphi_{lm,j}[\pi_{lm,j}]
\equiv \varphi_{lm}(r_j)[\pi_{lm,j}(r_j)] $} and {\small
$[\varphi_{lm,j}, \pi_{l'm',j'}] = i \delta_{l l'}\delta_{m
m'}\delta_{j j'}$}
(Note that the momenta $\pi_{lm,j}$ in Eq.(\ref{disc1}) are 
$a$ times the discretized versions of $\pi_{lm}$)	. 
This is of the form of the Hamiltonian of $N$ coupled HOs
(up to the overall factor of $1/a$, which does not change the entanglement
entropy to be computed) :
%
\be
H = \f{1}{2} \sum_{i=1}^N p_i^2 
+\f{1}{2} \sum_{i,j=1}^N x_i K_{ij} x_j ~.
\la{coupledham1}
\eeq
%
\hspace*{-8pt} 
with the interaction matrix elements
$K_{ij}$ given by ($i,j=1,\dots,N$):
\begin{widetext}
%
{\small
\beq
K_{ij} =  \frac{1}{i^2} 
\le[l(l+1)~\delta_{ij} + 
\f{9}{4}~\d_{i1} \d_{j1} 
+ \le( N - \f{1}{2}\ri)^2 \d_{iN} \d_{jN}
+ \le( 
\le(i+\f{1}{2}\ri)^2 + \le(i - \f{1}{2} \ri)^2
\ri)
\d_{i,j(i\neq 1,N)}
\ri] 
%
-\le[
\f{(j+\f{1}{2} )^2}{j(j+1)} 
\ri] \delta_{i,j+1}
-
\le[
\f{(i+\f{1}{2} )^2}{i(i+1)} 
\ri] \delta_{i,j-1}
\la{kij}
\end{equation}
}
A general eigenstate of the Hamiltonian (\ref{coupledham1}) is given by:
%
\be
\psi (x_1,\dots,x_N) = \prod_{i=1}^N N_i 
~H_{\n_i} \le( k_{D~i}^{\f{1}{4}}~{\underbar x}_i \ri)
\exp\le( -\f{1}{2} k_{D~i}^{\f{1}{2}}~{\underbar x}_i^2 \ri), \qquad
\mbox{where} \qquad 
N_i = \f{k_{D~i}^{\f{1}{4}}}{\p^{1/4}~\sqrt{2^{\n_i} \n_i!}}~.
\la{excwavefn1}
\ee
$ K_D \equiv U K U^T$ is a diagonal matrix ($U^TU=I_N$) with elements
$k_{Di}$, ${\underbar x} = Ux$, $\O = U^T K_D^{\f{1}{2}} U$, such that
$|\O| = |K_D|^{\f{1}{2}}$, $x^T = (x_1,\dots,x_N)$, ${\underbar x}^T =
({\underbar x}_1,\dots,{\underbar x}_N)$
and $\nu_i \, (i=1 \dots N)$ are indices of the Hermite polynomials
($H_{\nu}$). Note that the frequencies are ordered such that
$k_{Di}>k_{Dj}$ for $i>j$.

The density matrix, tracing over first $n$ of the $N$ field points, is given
by:
%
\ba
\rho \le( t; t'\ri) 
&&= 
\int \prod_{i=1}^n~dx_i~\psi (x_1,\dots,x_n;t_1,\dots,t_{N-n})~ 
\psi^\star (x_1,\dots,x_n;t_1',\dots,t_{N-n}') 
\nn \\
&&= \int \prod_{i=1}^n dx_i 
\exp\le[ -\f{x^T\cdot \O \cdot x}{2} \ri] 
\times
\prod_{i=1}^{N} N_i H_{\n_i} \le( k_{Di}^{\f{1}{4}} ~{\underbar x}_i\ri)
\times
{\exp\le[ -\f{x'^T\cdot \O \cdot x'}{2} \ri] 
\times
\prod_{j=1}^{N} N_j H_{\n_j} \le( k_{D~i}^{\f{1}{4}}~{\underbar x}'_i\ri)}
 \, ,
\la{denmatgen1}
\ea
\end{widetext}
where we have introduced the notation: $t_j \equiv
x_{n+j},~j=1..(N-n)$, i.e.  $x^T = (x_1,\dots,x_n;t_1,\dots,t_{N-1}) =
(x_1,\dots,x_n;t) $, where $t \equiv t_1,\dots,t_{N-n}$.  Density
matrix (\ref{denmatgen1}) yields the entanglement entropy via
(\ref{eq:def-vNE}). For an arbitrary excited state (\ref{excwavefn1}),
analytically, it is not possible to obtain a closed form expression
for $\rho(t;t')$. When all the HOs are in the GS however, i.e. when
$\n_i=0,~\forall i$, a closed form expression of $\rho(t;t')$ and the
corresponding entropy can be found \cite{bkls,sred}. Here, we will
investigate two non-trivial states, namely GCS and then ES.

Before we proceed with the evaluation of $S$, it is important to
compare and contrast between the two non-trivial states and the ground
state: (i) GCS, unlike GS, is not an energy eigenstate of HO. However,
GCS and GS are both minimum uncertainty states. (ii) ES, unlike GS,
are not minimum uncertainty states.
If the area law holds for both GCS and ES (or a superposition of ES),
then this would indicate that it is robust and unaffected by changes
of the chosen state. If it holds for GCS and not for ES (or a
superposition thereof), this might signal its validity {\it only} for
minimum uncertainty states; however, if the reverse, or some
other result holds true, a
simple interpretation cannot be given and more investigation would
have to be done.

First, let us assume that all the HOs are in GCS, i. e.,
%
\be
\psi_{_{GCS}} (x_1,\dots,x_N) = \le|\f{\O}{\p^N} \ri| 
\exp\le[- \sum_i \k_{Di}^{1/2} \le( {\underbar x}_i - \a_i \ri)^2\ri]
\la{cswf1}
\ee
%

\noindent where $\a^T=(\a_1,\dots,\a_N)$ represents the $N$ complex
GCS parameters $\a_i$ . Physically, the real and imaginary parts of
$\a_i$ corresponds to the classical position ($x_0$) and momentum
($p_0$) of the individual HOs respectively i. e., $\alpha_i = x_0 - i
p_0/k_{D i}$.
Defining,
\be
{\tilde x } \equiv  x - U^{-1}\a~,~~d\tilde x = dx \, ,
\la{cswf2}
\ee
it follows that 
{\small
\be
\psi_{_{GCS}} (x_1 \dots x_N) =
\le[\f{|\O|}{\p^N}\ri]^{\f{1}{4}} 
\exp\le[ -\f{{\tilde x}^T\cdot \O \cdot {\tilde x}}{2} \ri]
= \psi_0 ({\tilde x}_1 \dots {\tilde x}_N)~. 
\label{gswavefn}
\ee
}

\no where $\psi_0 ({\tilde x}_1 \dots {\tilde x}_N)$ is GS wave function.
Consequently, 
%
{\small
\ba
\rho_{_{GCS}} \le( t; t' \ri) 
&=& \int \prod_{i=1}^n dx_i
~\psi_{GCS} \le( x_i; t\ri) 
\psi_{GCS}^\star \le( x_i; t' \ri) \nn \\
&=& \int \prod_{i=1}^n d{\tilde x}_i
~\psi_{0} \le( {\tilde x_i} ; 
{\tilde t} \ri) 
\psi_{0}^\star \le( {\tilde x}_i;  {\tilde t}'\ri) 
\nn \\
&=& \le[ \f{|\O| }{|A| \p^{N-n}}\ri]^{\f{1}{4}} 
\exp\le[ -\f{1}{2} \le( t^T \gamma t + t'^T \gamma t' \ri)
+ t^T \beta t'  \ri] \nn \\
&=& \rho_0 (\tilde t; \tilde t'),   
\label{csmatrix1}
\ea
}
%
%
\noindent where 
\be
\O = \le( 
\begin{array}{cc} 
A & B \\
B^T & C 
\end{array}
\ri),
\ee
$A, B, C$ are $n\times n$, $n\times(N-n)$ and $(N-n)\times (N-n)$
matrices respectively, $\b = \f{1}{2} B^T A^{-1} B~$, $\gamma=C-\b ~$
and ${\tilde t} \equiv {\tilde t}_1, \dots, {\tilde t}_{N-n} = {\tilde
x}_n,\dots {\tilde x}_N$. Note that $B$, $B^T$ (and hence $\b$) are
zero if HOs are non-interacting.  Comparing with Eq. (11) of
Ref. \cite{sred}, we see that $\rho_0 (t,t')$ is precisely the GS
density matrix. That is, the GCS density matrix has the same functional
form, albeit in terms of the tilde variables.  Consequently, it will
have the same entropy as well, which is found in the following way.
By a series of transformations, (\ref{csmatrix1}) can be written as: 
%
\be 
\rho_{_{GCS}}(t,t') = 
\le[ \f{|\O| }{|A| \p^{N-n}}\ri]^{\f{1}{2}} \prod_{i=1}^{N-n} 
\exp\le[-\f{v_i^2+v_i'^2}{2} + {\bar \beta}_i v_i v_i' \ri] 
\la{keyabaat1}
\ee
%
\hspace*{-8pt} where $V\c V^T\equiv \c_D =$ diagonal, ${\bar\b} \equiv
\c_D^{-\f{1}{2}} V \b V^T \c_D^{-\f{1}{2}}$, $W {\bar \b} W^T\equiv
{\bar \b}_D= $ diagonal with elements ${\bar \b}_i$ and $v_i \in v
\equiv W^T (V \c V^T)^{\f{1}{2}} VT$. Since $\rho_{_{GCS}}$ in Eq. 
(\ref{keyabaat1}) is product of the $(N -n)$, two HO ($N=2, n = 1$)
density matrices, the total entropy is the sum of the entropies
\cite{sred}, i. e.,
%
\be
\! S = \sum_{i=1}^{N-n} 
- \ln[1-\xi_i] - \f{\xi_i}{1-\xi_i}\ln\xi_i \, , \quad 
\xi_i = \f{{\bar \b}_i}{1+ \sqrt{1-{\bar \beta}_i^2}} \, .
\la{ent2}
\ee
%
Since the eigenfunctions and eigenvalues are the same as those for the
ground state values, the total entropy is also same as that of the
ground state \cite{sred}:
%
\be
S = \sum_{l=0}^{l_{max}} (2 l+ 1) S_l = 
0.3 \le(\f{R}{a}\ri)^2 \, ,
\la{sredresult}
\ee
%
\hspace*{-8pt} 
where $R = a (n+ 1/2 )$, $S_l$ is the entropy for a given $l$.
Strictly speaking, the upper limit of the sum should have been
infinity, for which it is convergent.
Thus, we choose a maximum value of $l \equiv l_{max}$,  
such that $[S(l_{max}) -
S(l_{max} - 5)]/ S(l_{max} - 5) < 10^{-3}$ (The numerical error in
the total entropy is less than $0.1\%$). The sum in 
Eq.(\ref{sredresult}) is convergent. 
In other words, the entanglement entropy follows the area law,
even for {\it arbitrarily large} values of the coherent state
parameters $\a_i$!  Thus, one can conclude that all {\it classical
states} give rise to the area law. This is our first result.
It can be easily shown that the results continue to hold for the class
of SS, characterized by the same squeezing parameter $r$. The 
equivalents of Eqs.(\ref{cswf1}) and (\ref{cswf2}) are respectively: 
\begin{eqnarray}
\!\!\!\!\!
\psi_{_{SS}}(x_1,\dots,x_N) &=& \le|\f{\O}{\p^N} \ri| r^{N/2}
\exp\le[- \sum_i r \k_{Di}^{1/2} {\underbar x}_i^2\ri]\,,
\la{cswf1b} \\
{\tilde x } &\equiv&  \sqrt{r}~{\underbar x}~,~~d\tilde x = \sqrt{r}~
d{\underbar x}~.
\la{cswf2b}
\end{eqnarray}
Eqs.(\ref{gswavefn}) and (\ref{csmatrix1}) 
remain unchanged (up to irrelevant multiplicative factors), 
and so does the entropy (\ref{sredresult}).

Next, we assume the state to be in a linear superposition of $N$
wave-functions of the form in Eq.(\ref{excwavefn1}), such that in each
such wave-function there is {\it exactly one} HO in the first ES and
the rest $(N-1)$ in their GS.
%
%
%
Although non-trivial, we will find that this state is relatively easy to
handle. Using (\ref{excwavefn1}), the wave function for this state can be
simplified to
{\small
\ba
\psi_{_{1}}(x_1 \dots x_N)  
%
&=& 
\le| \f{\O}{4 \p^N}\ri|^{\f{1}{4}}
\sum_{i=1}^N 
a_i H_1 \le( k_{D i}^{\f{1}{4}} {\underbar x}_i\ri) 
\exp\le[ -\f{1}{2} \sum_{j} k_{D j}^{\f{1}{2}}~{\underbar x}_j^2 \ri], 
\nn \\
&=& \sqrt{2}~ 
\le( a^T K_D^{\f{1}{2}} {\underbar x} \ri)
 ~\psi_0\le( x_1,\dots,x_N, \ri)~,
\ea 
}
\noindent \hspace*{-8pt} 
where $a^T = \le( a_1,\dots, a_N \ri)$ are the expansion
coefficients (normalization of $\psi_1$ requires $a^T a=1$).

Substituting in Eq. (\ref{denmatgen1}), we get
%
\ba
\rho( t; t') =  2 \int \prod_{i=1}^n dx_i \, 
\le[{x}'^T \, \Lambda \, {x}^T  \ri] 
\,  \psi_0 \le(x_i; t \ri) \, \psi_0^\star \le(x_i; t'\ri) ,
\label{excden2}
\ea
%

\noindent where $\L$ is a $N \times N$ matrix and is defined as 
\beq
\L =  U^T~K_D^{\f{1}{4}}~a~a^T~K_D^{\f{1}{4}}~U \equiv
\le( 
\begin{array}{cc} 
\L_A & \L_B \\
\L_B^T & \L_C 
\end{array}
\ri) \, ,
\eeq
and $\L_A, \L_B, \L_C$ are $n\times n$, $n\times(N-n),(N-n)\times
(N-n)$ matrices respectively. Comparing Eqs. (\ref{csmatrix1},
\ref{excden2}), it is clear that the excited state density matrix is
{\it not} same as that of the ground state. More importantly, it is
{\it not} possible to obtain Eq. (\ref{csmatrix1}) from (\ref{excden2})
in any limit of $\L$. 
Integrating Eq. (\ref{excden2}), we get
\ba
\rho(t,t') &=& \rho_0(t,t') \,  Tr (\L_A A^{-1})  \nn \\
& \times & \le[1 -\f{1}{2} \le( 
t^T\L_\c t + t'^T \L_\c t' \ri) + t^T\L_\b t' \ri]\, , 
\label{excden3}
\ea
where we have defined
{\small
\ba
& &\hspace*{-13pt} 
\L_\gamma = \f{2\L_B^T \le(A^{-1} B \ri) - 
B^T \le(A^{-1}\ri)^T \L_A A^{-1} B}
{Tr(\L_A A^{-1})} \nn \\
& & \hspace*{-13pt} 
\L_\beta = \f{2\L_{_C} + B^T \le[A^{-1}\ri]^T \! \L_{_A} A^{-1} B 
- \L_{_B}^T A^{-1} B - B^T\!\le[A^{-1}\ri]^T\! \L_{_B}}
{Tr(\L_{_A} A^{-1})} \, .\nn 
\ea
}
Before proceeding further, we note the following: (i)
Eq. (\ref{excden3}) is the exact density matrix for the discretized
scalar field with any one HO in the first excited state while the rest
in the GS; (ii) unlike the GS (\ref{csmatrix1}), the excited state
density matrix (\ref{excden3}) contains non-exponential terms and
hence can not be written as a product of $(N-n)$, $2$-coupled HO
systems, as in Eq.(\ref{keyabaat1}).  Consequently, the entropy cannot
be written as a sum as in (\ref{ent2}).  However, for the vector $t^T$
outside the maximum of 
$t_{max}^T = \le(\f{3 (N-n)}{\sqrt{2
Tr(\c-\b)}} \ri)\le(1,1,\dots \ri)~,$
(corresponding to $3\sigma$ limits), the Gaussian  
inside $\rho_0(t;t')$ in (\ref{excden3}) is negligible. 
Thus, when the conditions 
\be
\e_1 \equiv t_{max}^T \, \L_{\b} \, t_{max} \ll 1 \quad 
\mbox{and} \quad 
\e_2 \equiv t_{max}^T \, \L_{\gamma} \, t_{max} \ll 1 \, 
\label{eq:condition}
\nn
\ee
are satisfied, one can make the approximation: 
%
\br
& & 1 -\f{1}{2} \le(t^T\L_\c t + t'^T \L_\c t' \ri) + t^T\L_\b t' \nn \\
& & \qquad \simeq 
\exp\l[-\f{1}{2} \le(t^T\L_\c t + t'^T \L_\c t' \ri) + t^T\L_\b t' \r]\, .
\label{eq:approx}
\er
%
\noindent Correspondingly, 
the density matrix (\ref{excden3}) takes the following simple form:
%
\ba
\rho(t,t') &=& \le[ \f{|\O| }{|A| \p^{N-n}}\ri]^{\f{1}{2}} 
Tr(\L_A A^{-1})  \nn \\
& \times& \exp\le(  -\f{1}{2} \le( t^T \gamma' t + t'^T \gamma' t' \ri)
+ t^T \beta' t'  \ri) \, ,
\la{almostsred1}
\ea
%
\noindent \hspace*{-8pt} 
where $\beta' \equiv \beta + {\L_\b}, \gamma' \equiv \gamma + {\L_\c}$.
Note that (\ref{almostsred1}) is of the same form as the ground state
density matrix (\ref{csmatrix1}), with matrices $\beta \rightarrow
\beta',\gamma \rightarrow \gamma'$ in the exponent, and up to
irrelevant normalization factors. The corresponding entropy will then
be given by Eq.(\ref{ent2}), with the replacements $\b \rightarrow
\b'$ and $\c\rightarrow \c'$ in the definition of $\xi_i$.  We tested
the validity of approximation (\ref{eq:condition}) numerically for
large values of $N$ ($N > 60$), using MATLAB \cite{nag}.  The error in
the approximation (\ref{eq:approx}) was less than $0.1\%$ for $a^{T} =
1/\sqrt{o}(0, \cdots 0, 1 \cdots 1)$ with the last $o$ columns being
non-zero.  The corresponding entropy, computed from the density matrix
(\ref{almostsred1}) was computed numerically, for $N=300$,
$n=100,\cdots,200$ and $o=10,20,30,40,50$.  

Before proceeding with the results we would like to mention the
following: For the above ES, the expectation value of energy is given
by:
\beq
{\cal E}_1 \equiv 
\langle	 H_{lm} \rangle 
=  {\cal E}_0 + \frac{1}{\sqrt{o}} \sum_{i=N-o+1}^{N} k_{Di}^{1/2}~,
\label{eq:ES-ener}
\eeq
where, $k_{Di}^{1/2}$ are the normal frequencies of the Hamiltonian
$H_{lm}$ and ${\cal E}_0$ is the ground state energy given by
\beq
{\cal E}_0 = \frac{1}{2} \sum_{i=1}^N k_{Di}^{1/2} \, .
\label{eq:GS-ener}
\eeq
Note that we have not set the GS energy to zero and, as mentioned
earlier, the frequencies are ordered such that $k_{D_i} > k_{D_j}$ for
$i > j$. Rewriting Eq. (\ref{eq:ES-ener}), we have:
\beq
{\cal E}_1
=
\frac{1}{2} \sum_{i=1}^{N-o} k_{Di}^{1/2} 
+ \left(\frac{1}{2} + \frac{1}{\sqrt{o}} \right)
\sum_{i=N-o+1}^N k_{Di}^{1/2}~.
\la{energy10}
\eeq
%
%
%
%
%
As mentioned before, the $k_{Di}$ are in ascending order. 
Moreover, the last few terms in the second sum in
Eq.(\ref{energy10}) dominate.
Consequently for $N=300$ and $o=10 \dots 50$, $({\cal E}_1 -
{\cal E}_0)/{\cal E}_0 \approx 0.3 \dots 0.6$ i.e.  the energy of
these ES are about $30-60\%$ higher than the GS energy. Note that
these energies are in units of $1/a$, with $a$ the ultraviolet cut-off
(the lattice spacing).  Thus, if we choose the latter to be of the
order of the Planck length, as is reasonable in any theory of quantum
gravity, the GS and ES energies (as well as the energy density of the  
ES) are Planckian, where we have ignored
the resulting gravitational self-interactions of the system.


%
\begin{figure}[!htb]
\begin{center}
\epsfxsize 3.00 in
\epsfysize 2.50 in
\epsfbox{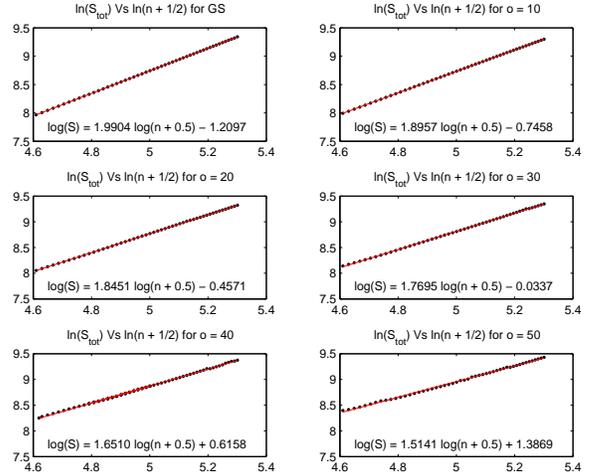}
\caption{Logarithm of GS and ES entropies versus the
radius of the sphere ($R/a$) i. e. $R = a (n + 1/2)$ for $N = 300$ and
$100 \leq n \leq 200$.  We choose the maximum value of $l$ such that
$[S(l_{max}) - S(l_{max} - 5)]/ S(l_{max} - 5) < 10^{-3}$.  The
numerical error in the total entropy is less than $0.1\%$.}
\label{fig:1}
\end{center}
\vspace*{-0.05cm}
\end{figure}
%
%
\begin{figure}
\begin{center}
\epsfxsize 3.00 in
\epsfysize 2.50 in
\epsfbox{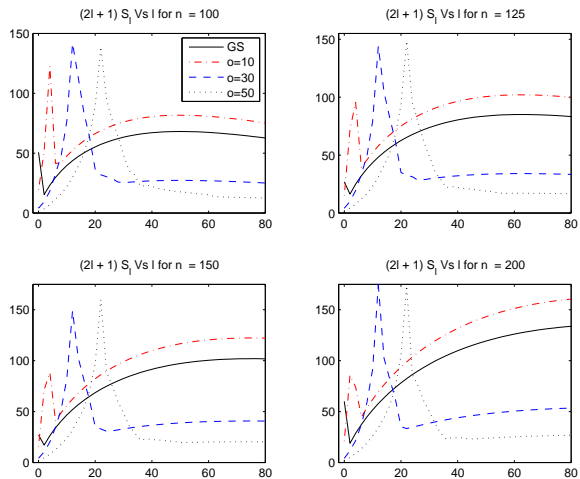}
\caption{Plot of the distribution of entropy per partial wave $[(2 1 +
    1) S_l]$ for GS (solid-curves) and ES (dotted-curves). To
  illustrate the difference between the GS and ES (and that all curves
  can be fitted in the same graph), we have multiplied the GS entropy
  per partial wave by a factor of 5, while the $o=10$ and $o=30$
  curves have been multiplied by factors of 6 and 2 respectively in
  each plots.}
\label{fig:2}
\end{center}
\vspace*{-0.55cm}
\end{figure}
In Fig.(1), we have plotted $\log(S)$ versus $\log(R/a)$. From the
best-fit curves, we see that for $o=10,20,30,40,50$, 
$S=0.4744(R/a)^{0.9479}$, $0.6331(R/a)^{0.9223}$, $0.9669(R/a)^{0.8848}$,
$1.8511	(R/a)^{0.8255}$, $4.002(R/a)^{0.7571}$ respectively. Thus,
although the coefficient in front increases, the power decreases with
the number of excited states, and for large enough areas, the GS 
(or closely related GCS or SS) entropy is greater than 
the ES entropy. 
We would thus like to conjecture that if the entanglement entropy 
of a superposition of the GS and ES is computed, it would (at least
approximately) be a sum
of the GS entropy (the area law) and the ES state entropy that we 
found, in which case, the latter can be interpreted as 
(power-law) corrections to the area law. Such corrections can be contrasted
with entropy corrections obtained from other sources
\cite{dmb}.  
%
%
%
In Fig.(2), we have plotted the entropy for each partial wave, 
$(2l + 1) S_l$, versus $l$, for $N=300$ and 
various values of $n$. For each $n$, we have plotted for 
$o=10,30,50$.
It can be seen that for the GS, there is a maxima at $l=0$, after
which $(2l+1)S_l$ decreases. Once it reaches a minimum, it starts to
rise again, due to the large degeneracy factor $(2l+1)$. For the ES
however, a sharp maximum occurs between $l=5$ and $l=30$, depending on
the parameter $o$.  We hope to get a better understanding of this
phenomenon in the future.

%

To summarize, in this work we have computed the entanglement entropy
of scalar fields, after tracing over its degrees of freedom inside a
hypothetical sphere of radius $R$. The oscillator modes representing
these degrees of freedom were assumed to be in GCS, SS 
and a superposition of the first ES. In the case of GCS and SS, 
the entropy turned
out to be identical to that for the GS of the form $0.3 (R/a)^2$,
while for the ES, the entropy goes as a power of the area which is
less than unity.  
%
%

In the light of the above results, let us discuss the implications of
our results to the BH entropy: the Bekenstein-Hawking area law not
only tells that the BH entropy is proportional to area, it also gives
the precise value of the proportionality constant [$1/(4
\ell_{_P}^2)$]. Our analysis, for the MST, suggests that the constant of
proportionality {\it as well} as the power of the area 
depend on the choice of the state of the scalar
field. This raises an immediate question: which states determine 
the Bekenstein-Hawking entropy? \cite{hsu,DS-new}.
As stated before, it appears that the GS is most 
relevant, at least for large areas, while the ES give rise to
correction terms. 	

Open problems include extending our analysis to higher ESs.  One
technical problem that we anticipate in this case is that the density
matrix will not be expressible as the GS matrix with shifted
parameters (such was the case for (\ref{excden3})), since $H_n(x) \sim
x^n $.  Finally, analytical proofs of the area law, for GCS, SS and ES, 
which do not depend on the shape of the traced out volume, 
along the lines of Ref. \cite{eisert} would be
illuminating. We hope to examine these issues elsewhere.

We would like to thank M. Ahmadi, R. K. Bhaduri, C. Burgess,
A. Dasgupta, J. Gegenberg, A. Ghosh, V. Husain, G. Kunstatter, S. Nag,
A. Roy, T. Sarkar and R. Sorkin for useful discussions. We thank
J. Eisert, C. Kiefer and L. Sriramkumar for comments on the earlier
version of the draft. We also thank the anonymous referees for several
important comments and suggestions. SS would like to thank the
Dep. of Physics, Univ. of Lethbridge for hospitality where most of
this work was done. This work was supported in part by the Natural
Sciences and Engineering Research Council of Canada.

\end{document}